# Effect of interface resistance on thermoelectric properties in (1-$x$)La$_{0.95}$Sr$_{0.05}$Co$_{0.95}$Mn$_{0.05}$O$_3$ /($x$)WC composite

Ashutosh Kumar[1,*], Krzysztof T. Wojciechowski[2]

[1]Lukasiewicz Research Network - Krakow Institute of Technology, Zakopianska 73, 30-418 Krakow, Poland

[2]Faculty of Materials Science and Ceramics, AGH University of Science and Technology, 30-059 Krakow, Poland

**Abstract:** In this study, the synergistic effect of the particle size of the dispersed phase and the interface thermal resistance ($R_{int}$) between the phases on the phonon thermal conductivity ($\kappa_{ph}$) of the (1-$x$)La$_{0.95}$Sr$_{0.05}$Co$_{0.95}$Mn$_{0.05}$O$_3$/($x$)WC thermoelectric composite is demonstrated. Further, the correlation between the $R_{int}$ and the Kapitza radius is discussed using Bruggeman's asymmetrical model. In particular, the synthesis of polycrystalline La$_{0.95}$Sr$_{0.05}$Co$_{0.95}$Mn$_{0.05}$O$_3$ is performed using a standard-solid state route. The presence of WC nanoparticles is confirmed from the electron microscopy images. Electrical conductivity ($\sigma$) increases, and the Seebeck coefficient ($\alpha$) decreases with the increase in conducting WC volume fraction in the composite. The simultaneous increase in $\sigma$ and a decrease in $\kappa_{ph}$ with the WC volume fraction results in an increased figure of merit (zT) for (1-$x$)La$_{0.95}$Sr$_{0.05}$Co$_{0.95}$Mn$_{0.05}$O$_3$/(x)WC composite. A maximum zT ~0.20 is obtained at 463 K for (1-$x$)La$_{0.95}$Sr$_{0.05}$Co$_{0.95}$Mn$_{0.05}$O$_3$/($x$)WC composite with $x$=0.010. This study shows promise to design thermoelectric composites with the desired phonon thermal conductivity considering the elastic properties between the phases.

[#]**Email: ashutosh.kumar@kit.lukasiewicz.gov.pl**

# INTRODUCTION

Wastage of heat energy from several energy sources while in practical use has been a concern for decades. [1] Thermoelectric (TE) devices are promising in waste heat retrieving technology for the efficient use of available energy resources.[2,3] The usefulness of a TE device is estimated using a dimensionless quantity known as a figure of merit (zT= $\sigma\alpha^2 T/(\kappa_{ph}+\kappa_e)$), where $\alpha$ is the Seebeck coefficient, $\sigma$ is the electrical conductivity, T is the absolute temperature, and $\kappa$ (=$\kappa_{ph}+\kappa_e$) is the total thermal conductivity, and consists of phonon thermal conductivity ($\kappa_{ph}$) and electronic thermal conductivity ($\kappa_e$). The interdependence between $\sigma$, $\alpha$, and $\kappa_e$ is challenging to achieve a high zT in a single material.[4] The increase in $\sigma$ reduces $\alpha$; also, an increase in $\sigma$ increases $\kappa_e$, according to Wiedemann Franz law. Ioffe suggested that a low value of $\kappa_{ph}$ is essential for achieving a high zT in TE materials. The reduction in $\kappa_{ph}$ has been demonstrated in literature through several strategies, including lattice defects[5], mass disorder[6], nanostructuring[7], artificial superlattices[8], and or preparation of composite materials[9] that enhances the scattering of phonons. Nevertheless, the scattering of phonon through lattice defects, large lattice anharmonicity, and impurities in a material can also scatter electrons, leading to reduced electrical conductivity[4]. Moreover, making composite materials can be promising for lowering $\kappa_{ph}$, considering the mismatched elastic properties (sound velocity) and the Kapitza radius between the phases.[10]

Oxide materials are exciting candidates for energy conversion technologies for power generation at higher temperatures owing to their chemical and thermal inertness in an open environment.[3] Among the oxide materials, cobalt-based systems ($LaCoO_3$ [11–13], $Ca_3Co_4O_9$ [14,15], $Na_xCoO_2$ [16,17]) are promising for TE applications because of the existence of several charges states and corresponding different spin states of cobalt.[13,18] However, these oxide materials possess poor zT due to low $\sigma$ and high $\kappa_{ph}$.[19] An improvement in zT is obtained in several oxide composites, viz. $Ca_3Co_4O_9$/Ag [20], $LaCoO_3$/graphene[21], $LaCoO_3$/$La_{0.7}Sr_{0.3}MnO_3$ [22,23], due to reduced $\kappa_{ph}$. A notable reduction in $\kappa_{ph}$ was also shown in several composites with the addition of a nanostructured secondary phase[24–26]. However, the lowering of $\kappa_{ph}$ in nanostructured composite materials is mainly ascribed to the quantum size effects [19,27]. Also, these studies are often focused on general transport properties and do not consider the impact of interface thermal resistance ($R_{int}$) and acoustic impedance mismatch (AIM) between the phases on $\kappa_{ph}$. On the other hand, these parameters are used in the analysis of heat transport mechanisms in several ceramic (ZnS/diamond [10], SiC/Al [28], ZnO/In$_2$O$_3$ [29]) and polymer composites (glass/epoxy [9,30]). Moreover, this method did not receive significant research attention in optimizing the thermal conductivity for TE composite materials. In our recent article, reduction in the thermal conductivity of the composite, even lower than the thermal conductivity of the matrix is obtained, when the particle size of the dispersed phase is smaller than the Kapitza radius [31]. However, the electrical conductivity of the dispersed phase is not promising and hence requires additional research to improve the TE properties in composite systems.

Because of the gap in the literature as mentioned above, a systematic study on the TE properties of a composite oxide that consists of La$_{0.95}$Sr$_{0.05}$Co$_{0.95}$Mn$_{0.05}$O$_3$ (a promising cobalt-based oxide system, zT=0.14 at 480 K [11]) and highly conducting WC (with different elastic properties than LSCMO) is performed. The addition of conducting WC improves the electrical conductivity, and different elastic properties between the phase reduce $\kappa_{ph,}$ and hence suitable for improved TE properties. In particular, the role of $R_{int}$ between the phases and particle size of the dispersed phase on $\kappa_{ph}$ is analyzed. Further, the experimental results for thermal conductivity are discussed using Bruggeman's asymmetrical model.

## EXPERIMENTAL SECTION

The standard solid-state route was employed to synthesize the polycrystalline La$_{0.95}$Sr$_{0.05}$Co$_{0.95}$Mn$_{0.05}$O$_3$ (LSCMO) sample, as shown in our previous studies.[11] The (1-$x$)La$_{0.95}$Sr$_{0.05}$Co$_{0.95}$Mn$_{0.05}$O$_3$/($x$)WC composite with $x$=0.000, 0.002, 0.005, 0.010, 0.020, and 0.050 is prepared by mixing the different volume fraction ($x$) of WC nanoparticle (Sigma Aldrich, 150-200 nm). The composite mixture was further ground to mix homogeneously and then sintered using a spark plasma sintering (SPS) technique at 1023 K in a uniaxial pressure of 60 MPa. The obtained cylindrical pellets of 10 mm diameter and 12-13 mm height were cut to proper dimensions using a precise wire saw for further measurements. The microstructure investigation of the polished sample surface was performed using a scanning electron microscope. X-ray diffraction of samples was obtained by the Rigaku diffractometer ($\lambda$=1.5406 Å). The surface mapping of the Seebeck coefficient for the (1-$x$)LSCMO/($x$)WC composite was done using a scanning thermoelectric microprobe (STM) at 298 K with a spatial resolution of 50 μm. Thermal conductivity ($\kappa$=Dc$_p\rho$) of the samples was estimated by measuring the thermal diffusivity (D) using laser flash analysis (LFA-457, NETSCH) apparatus in the Ar (5N) atmosphere (30 ml/min) with an uncertainty of 7%. Specific heat ($c_p$) was determined from the Dulong-Petit law. The bulk density ($\rho$) of the sample was measured using its mass and geometric volume. Electrical conductivity and Seebeck coefficient were measured under Ar (5N) atmosphere (50 ml/min) by the SBA 458 (NETSCH) apparatus. The uncertainty of the measurement of $\alpha$ and $\sigma$ is 7% and 5%, respectively.

## RESULTS AND DISCUSSION

The structural analysis of the $(1\text{-}x)\text{La}_{0.95}\text{Sr}_{0.05}\text{Co}_{0.95}\text{Mn}_{0.05}\text{O}_3/(x)\text{WC}$ composite is performed using the x-ray diffraction method, and the corresponding diffraction pattern is shown in Fig. 1(a). The LSCMO shows a characteristic peak at $2\theta = 32.74°$ (104) and $32.88°$ (110) having the following lattice parameters (a=b=5.443 Å, c=13.154 Å). In composite samples, only intensity due to LSCMO is observed without any trace of impurity within the sensitivity of XRD. Also, the reflections due to WC is not prominently observed and may be ascribed to its lower volume fraction in the composite. The diffraction pattern corresponding to WC used in the composite is also shown in Fig.1.

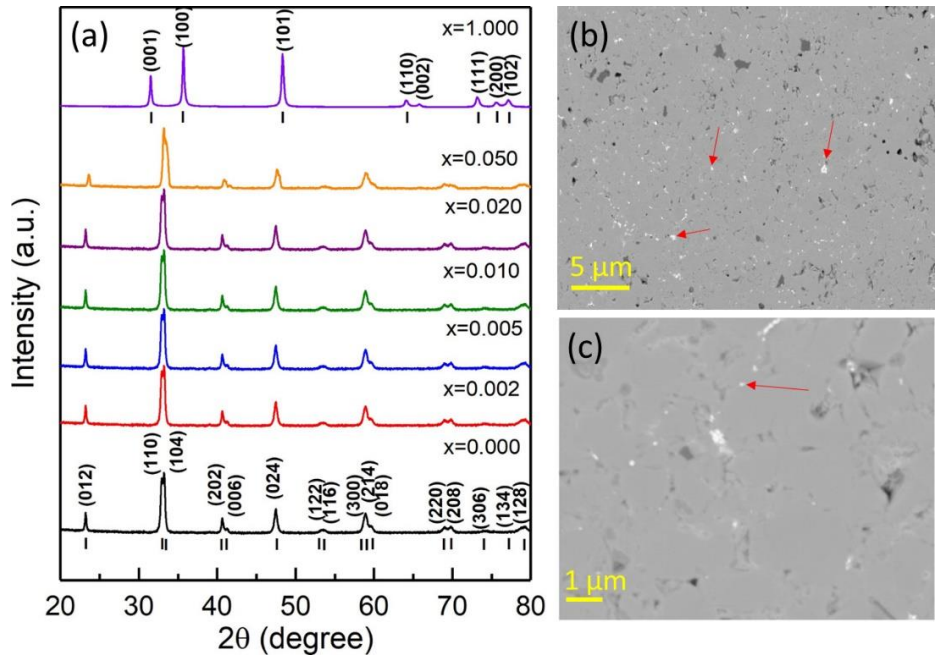

Figure 1: (a) X-ray diffraction pattern for $(1\text{-}x)\text{La}_{0.95}\text{Sr}_{0.05}\text{Co}_{0.95}\text{Mn}_{0.05}\text{O}_3/(x)\text{WC}$ composite. Bragg's position and Miller indices for $\text{La}_{0.95}\text{Sr}_{0.05}\text{Co}_{0.95}\text{Mn}_{0.05}\text{O}_3$ and WC are marked. Scanning electron miscroscopy (SEM) images for $(1\text{-}x)\text{La}_{0.95}\text{Sr}_{0.05}\text{Co}_{0.95}\text{Mn}_{0.05}\text{O}_3/(x)\text{WC}$ composite for (b, c) $x$=0.010.

The surface morphology of the polished surface of (1-*x*)LSCMO/(*x*)WC composite for *x*=0.010 is shown in Fig. 1(b-c). The presence of WC nanoparticles (shown by red arrows) is seen in the composite samples. The WC nanoparticles are at the grain boundaries. The WC particle size lies between 150-200 nm, and LSCMO particles are in the range of few micrometers (1-3 μm). The bulk and theoretical densities of the composite were used to find the relative density. The theoretical densities for the composite samples are calculated through the volume-weighted arithmetic mean of each phase in the composite. The relative density for all the samples is greater than 95%.

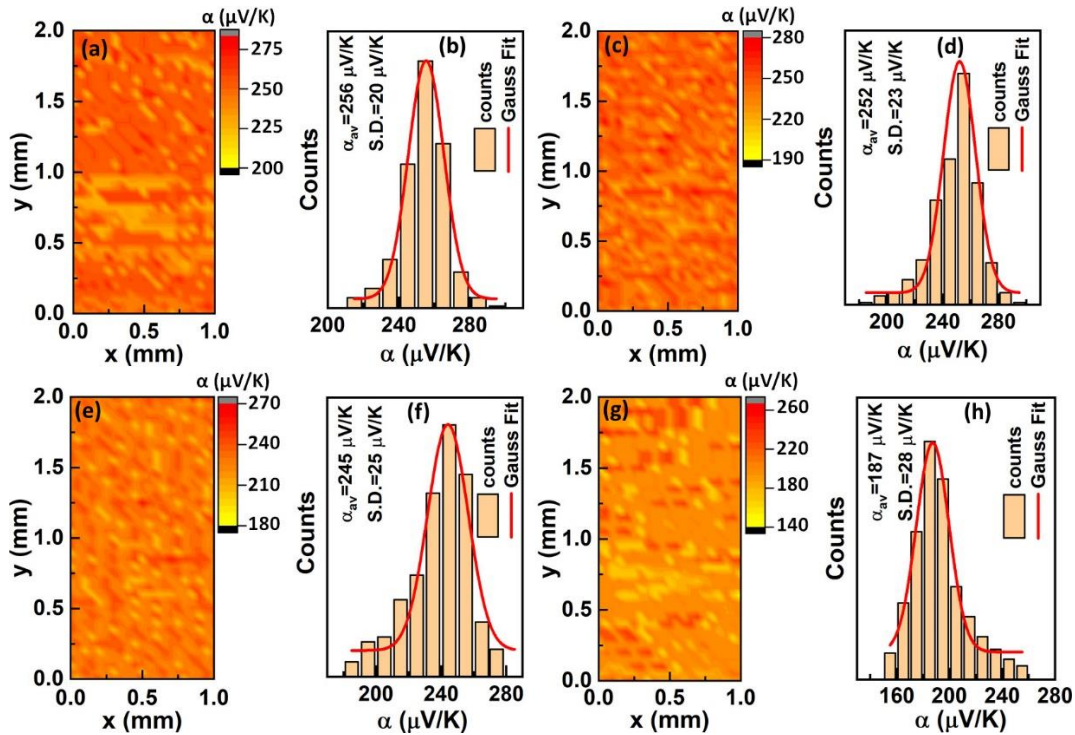

Figure 2. Spatial mapping of Seebeck coefficient using scanning thermoelectric microprobe (STM) and corresponding distribution curve for (1-*x*)La$_{0.95}$Sr$_{0.05}$Co$_{0.95}$Mn$_{0.05}$O$_3$/(*x*)WC composite (a-b) *x*=0.000, (c-d) *x*=0.005, (e-f) *x*=0.010, and (g-h) *x*=0.050.

The spatial mapping of the Seebeck coefficient (α) on the surface of the composite samples, measured using a scanning thermoelectric microprobe at 300 K, and the corresponding distribution of Seebeck coefficient is shown in Fig. 2(a-h). The average Seebeck coefficient for *x*=0.000 is 256 μV/K and decreases to 245 μV/K for *x*=0.010 to 187 μV/K for *x*=0.050. It suggests an increase in the electrical conductivity of the composite and consistent with the earlier studies.[32] It is noted that although there are two phases in the composite, single model distribution in the histogram curve for the composite samples is observed. In other words, the Seebeck coefficient corresponding to WC only (~5-10 μV/K) is not observed in any of the composite samples. It is corroborated to the lower STM tip resolution (50 μm) as well as, the higher diameter of the tip (in few micrometers) compared to the particle size of WC (100-150 nm). However, the average value of the Seebeck coefficient decreases with an increase in the WC volume fraction ($x$).[21] It is attributed to the collective contributions from both LSCMO and WC phases.

The temperature-dependent Seebeck coefficient (α) for the LSCMO/WC composite is shown in Fig. 3(a). The positive sign of α suggests a p-type dominating charge carrier in the system.[11] At 300 K, α for LSCMO is 260 μV/K. For x=0.002, α increases to 263 μV/K and decreases with the further increase in $x$ in the composite. The Seebeck coefficient decreases from 260 μV/K for x=0.000 to 255 μV/K for x=0.005 to 251 μV/K for x=0.010 to 231 μV/K for x=0.020 to 186 μV/K for x=0.050. A decrease in α with the increase in temperature is observed for all the samples. Also, the slope change in the α vs. temperature curve is identical for all samples, depicting that the transition in the spin-states is identical in all the samples. [16,33,34]

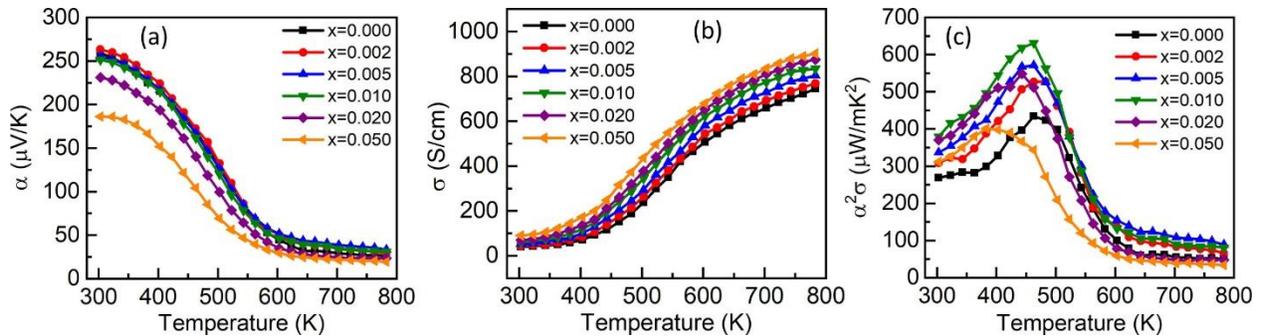

Figure 3(a) Seebeck coefficient (α), (b) Electrical conductivity (σ), and (c) power factor (α²σ) as a function of temperature for (1-$x$)La$_{0.95}$Sr$_{0.05}$Co$_{0.95}$Mn$_{0.05}$O$_3$/($x$)WC composite.

Figure 3(b) shows the temperature-dependent electrical conductivity (σ) for all the composite samples. The σ increases with an increase in $x$ in the composite and is attributed to the significantly high σ of WC (50000 S/cm @ 300 K).[35] The increase in σ with $x$ in the composite is consistent with the decrease in α. Also, the rise in σ with increasing temperature for all the samples dictates the semiconducting behavior. The σ for LSCMO at 300 K is ~40 S/cm, and it increases to ~60 S/cm for $x$=0.010 and to ~90 S/cm for $x$=0.050. A small rise in σ in the composite with WC, having too high electrical conductivity (50000 S/cm), may be attributed to the presence of contact resistance between the LSCMO and WC phases. Further, the power factor (α²σ) was calculated using the measured values of σ and α at each temperature for all the samples and is shown in Fig. 3(c). The α²σ increases with $x$ and is due to the rise in σ with a minimal reduction in α. It is also noted that α²σ first increases and decreases with the temperature rise and is in line with the previous studies in cobaltates.[11,23] The decrease in power factor at higher temperatures is attributed to a decrease in the Seebeck coefficient.

The temperature-dependent total thermal conductivity (κ) of the LSCMO/WC composite is shown in Fig.4(a). The κ increases with a rise in temperature for all samples and is attributed to an increase in σ. Also, κ of the composite increases with the increase in $x$. This increase in κ may be attributed to the high κ of the WC [36]. However, it is noted that the rise in κ for the composite samples does not follow the rule of mixture, i.e., the increase in κ is non-monotonous. Hence, to understand the exact nature of thermal conductivity in the composite samples, electronic (κ$_e$) and phonon thermal conductivity (κ$_{ph}$) was separated from κ. The κ$_e$ is calculated using the Wiedemann-Franz law, i.e., κ$_e$=LσT, where L is the Lorenz number and is calculated using the following equation L=1.5+exp(-|α|/116) [37].

The κ$_{ph}$ is obtained by subtracting the κ$_e$ from κ. Temperature-dependent κ$_{ph}$ for the composites is shown in Fig. 4(b). The κ$_{ph}$ increases with the rise in temperature up to ~500 K and decreases with a further temperature increase. The decrease in κ$_{ph}$ at higher temperatures is attributed to the decrease in the phonon mean free path that enhances the phonon scattering. However, κ$_{ph}$ as a function of $x$ decreases with the increase in $x$ in the composite. To elucidate the decrease in κ$_{ph}$ with the increase in $x$ in the composite, the acoustic impedance mismatch (AIM) model and the Debye model are used to calculate the interface thermal resistance ($R_{int}$) between the phases. Further, Bruggeman asymmetrical model, which considers the $R_{int}$, was used to estimate the theoretical κ$_{ph}$ for the composites and compared with the experimental values.

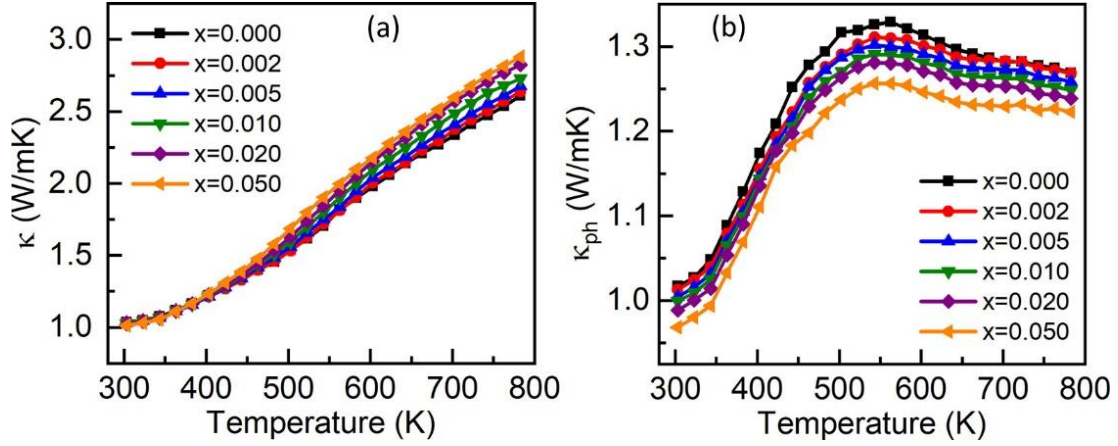

Figure 4 (a) Total thermal conductivity ($\kappa$), (b) phonon thermal conductivity ($\kappa_{ph}$) as a function of temperature for $(1-x)La_{0.95}Sr_{0.05}Co_{0.95}Mn_{0.05}O_3/(x)WC$ composite.

When the phonons travel in a composite material meet the interface between the phases, the probability of phonon reflected and transmitted at the interface can be estimated using the acoustic impedance mismatch (AIM) model. First, the fraction of phonons ($q$) with the angle of incidence at the interface between two phases is estimated using the measured sound velocities of the matrix ($v_m$) and the dispersed phase ($v_d$) using the following equation: $q = \frac{1}{2}\left(\frac{v_m}{v_d}\right)^2$ [10], The measured sound velocities and acoustic impedances for LSCMO and WC are shown in Table I, and it gives a $q$=0.128. This means that approximately 12.8% of the phonons incident at the interface between the phases within the critical angle. Next, the probability of phonon transmission ($p$) at the interface between LSCMO and WC is calculated using the following equation: $p = \frac{4Z_m Z_d}{(Z_m + Z_d)^2}$ [38], where $Z_i = v_i\rho_i$ is called the acoustic impedance, $v_i$ is the velocity of the $i^{th}$ material. The $Z_i$ calculated for both LSCMO ($Z_m$) and WC ($Z_d$) results to $p$ value of 0.581. This indicates that the phonons with incident angle within the critical angle, 58.1 % of them may transmit through the interface. Hence, the transmission probability ($\eta = pq$) for these LSCMO/WC interface is 0.0746. The schematic showing the probability of transmission ($p$) and reflection (1-$p$) of phonons at the interface between two materials in a composite vs. acoustic impedance mismatch ($Z_A/Z_B$) is depicted in Fig. 5(a). However, according to the AIM model, for considerable phonon reflection (1- $p$), $Z_A/Z_B$ should be too high for solid samples. A dotted line marks the mismatch in the acoustic impedance between the LSCMO and WC phase. As can be seen, this mismatch is not sufficient to scatter all the phonons to reduce the $\kappa_{ph}$ of the composite.

TABLE I: The experimental bulk density ($\rho$), measured sound velocity ($v$) in transverse ($v_t$) and longitudinal ($v_l$) directions, corresponding acoustic impedance ($Z$), and the coefficient of phonon transmission at the interface (p) for LSCMO and WC samples.

| Sample Name | $\rho$ (g/cm$^3$) | $v$ (m/sec) | | Z (kg/(m$^2$s)) | | $p$ (%) |
|---|---|---|---|---|---|---|
| | | $v_t$ (Trans.) | $v_l$ (Long.) | Trans. | Long. | |
| LSCMO | 5.94 | 1600 | 2610 | 13246 | 25066 | 58.1 |
| WC | 15.43 | 2110 | 3170 | 67892 | 110787 | |

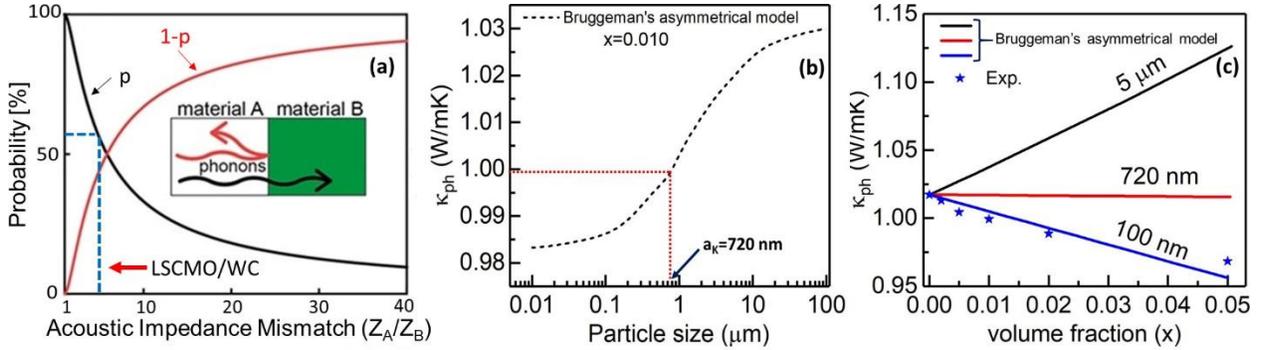

Figure 5: (a) A scheme depicting the change in the probability for phonon transmission (p) and reflection (1-p) as a function of acoustic impedance ratio ($Z_A/Z_B$) between two materials A and B. The $Z_A/Z_B$ is marked for the LSCMO/WC composite. (b) The calculated phonon thermal conductivity ($\kappa_{ph}$) of (1-$x$) La$_{0.95}$Sr$_{0.05}$Co$_{0.95}$Mn$_{0.05}$O$_3$/($x$)WC composite for $x$=0.010 as a function of different particle size of WC at 300 K, obtained from the Bruggeman's asymmetrical model. (c) Calculated $\kappa_{ph}$ as a function of WC volume fraction ($x$) for the different particle size of WC at 300 K. The symbols represent the $\kappa_{ph}$ obtained from total thermal conductivity.

There is a strong correlation between the interface thermal resistance ($R_{int}$) between the phases of the composite materials and the particle size of the dispersed phase. Hence, $\kappa_{ph}$ of a composite can be reduced below the $\kappa_{ph}$ of the matrix if the particle size of the dispersed phase is smaller than the Kapitza radius (a$_K$) between the phases [31]. Therefore, we use the Debye model to calculate the $R_{int}$ for the LSCMO-WC phase by the equation: $R_{int} = \frac{4}{\rho c_p v_D \eta}$ , where $c_p$ is the specific heat capacity, $v_D \left( = \left( \frac{3}{\frac{1}{v_l^3} + \frac{2}{v_t^3}} \right)^{1/3} \right)$ is the Debye velocity, calculated using the measured sound velocities in longitudinal ($v_l$) and transverse ($v_t$) directions. The $R_{int}$ between the LSCMO and WC phases is $7.05 \times 10^{-7}$ m$^2$K/W. Further, a$_K$ (=$R_{int}.\kappa_m$) for the LSCMO/WC composite is ~720 nm at 300 K, where $\kappa_m$ is the phonon thermal conductivity of the matrix (LSCMO). Next the correlation between the $R_{int}$ and $a_K$ on $\kappa_{ph}$ of the composite is depicted using the Bruggeman's asymmetrical model, given by the formula [39]

$$(1-x)^3 = \left(\frac{k_m}{k}\right)^{\frac{1+2\alpha}{1-\alpha}} \left(\frac{k-k_d(1-\alpha)}{k_m-k_d(1-\alpha)}\right)^{\frac{3}{1-\alpha}}$$

Using the phonon thermal conductivity of dispersed phase ($\kappa_d$), and the matrix ($\kappa_m$) with $\alpha = a/a_K$, where $a$ is defined as the particle size of the dispersed phase; the phonon thermal conductivity of the composite can be estimated. Figure 5(b) depicts the change in $\kappa_{ph}$ of the composite for $x$=0.010 with different particle sizes ($a$). It is noted that with the decrease in particle size, the $\kappa_{ph}$ decreases. It suggests that the $\kappa_{ph}$ of the composite can be reduced with the constant value of $R_{int}$, if the particle size of dispersed phases is reduced. For particle size smaller than the Kapitza radius, $\kappa_{ph}$ of the composite is smaller than that of the matrix. It confirms the strong correlation between $R_{int}$ and $a_K$. The $a_K$ value for the present case is marked in Fig.5(b).

The change in $\kappa_{ph}$ as a function of $x$ for three different particle size of WC ($>a_K$, $= a_K$, $< a_K$) is shown in Fig. 5(c). It is observed that $\kappa_{ph}$ of the composites increases with $x$, for $a > a_K$ (~720 nm). It suggests that even there is an acoustic impedance mismatch between the LSCMO and WC, the $\kappa_{ph}$ does not decrease for $a > a_K$ and hence further confirms the correlation between $R_{int}$ and $a_K$ [10,31].

Further, an almost constant value of $\kappa_{ph}$ is observed for $a \sim a_K$. However, $\kappa_{ph}$ of the composite gets reduced for $a < a_K$. The experimental value of the $\kappa_{ph}$ (star symbol) agrees well to the theoretical prediction (solid lines). The decrease in $\kappa_{ph}$ for $a < a_K$ is attributed to the increase in the contact surface between LSCMO and WC phases because of the high surface to volume ratio caused by small particle size [31]. Hence, the $R_{int}$ between the LSCMO and WC phases enhances the phonon scattering when $a < a_K$ and hence reduces $\kappa_{ph}$.

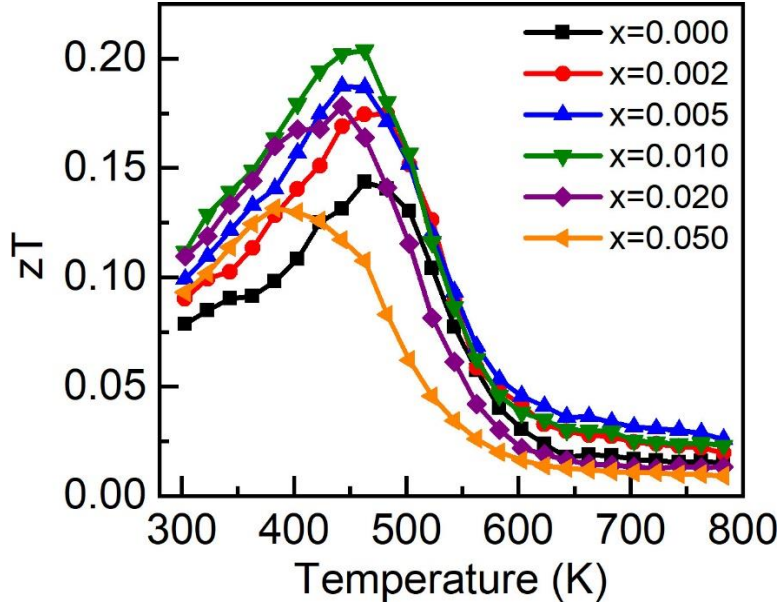

FIG. 6: (a) Figure of merit (zT) as a function of temperature for (1-$x$)La$_{0.95}$Sr$_{0.05}$Co$_{0.95}$Mn$_{0.05}$O$_3$/($x$)WC composite.

The experimental values of $\alpha$, $\sigma$, and $\kappa$ are utilized to calculate the thermoelectric figure of merit (zT) for the LSCMO/WC composite. Figure 6 depicts the temperature-dependent zT for the LSCMO/WC composite. The zT enhances at 300 K with $x$ and is due to increased $\sigma$ and reduced $\kappa_{ph}$. An increase in zT with temperature is also obtained up to 463 K. The maximum zT for LSCMO/WC composite is 0.20 at 463 K, one of the highest zT obtained in the LaCoO$_3$ based system at this temperature [40–42].

**CONCLUSION**

In this report, we show the correlation between the interface thermal resistance and particle size ($a$) of the dispersed phase on the phonon thermal conductivity ($\kappa_{ph}$), and a reduced $\kappa_{ph}$ is obtained for the LSCMO/WC composite for $a$ smaller than the Kapitza radius ($a_K$), using the Bruggeman's asymmetrical model. In particular, a standard solid-state reaction is used to synthesize the polycrystalline LSCMO. The pure phase formation of LSCMO and the presence of WC in the composite was confirmed from the x-ray diffraction and scanning electron microscopy analysis, respectively. The addition of WC (having mismatched elastic properties than LSCMO) improves the electrical conductivity ($\sigma$) with a minimal reduction in the Seebeck coefficient. The phonon thermal conductivity ($\kappa_{ph}$) decreases with the increase in the WC volume fraction and is analyzed using the acoustic impedance mismatch model and the Bruggeman asymmetrical model. The $\kappa_{ph}$ of the composite decreases for $a<a_K$. The simultaneous increase in $\sigma$ and decrease in $\kappa_{ph}$ with the addition of WC nanoparticles improve the figure of merit (zT) in the system. A maximum zT of 0.20 is obtained for LSCMO/WC composite with x=0.010 at 463 K. The results obtained in this study can generate excitement in composite materials to optimize the electrical and thermal conductivity based on the elastic properties of materials for improved TE applications.

## ACKNOWLEDGMENTS

The authors would like to thank the Foundation for Polish Science for financial support (TEAM-TECH/2016-2/14 grant "New approach for the development of efficient materials for direct conversion of heat into electricity") under the European Regional Development Fund. The beneficiary institution of the grant is The Lukasiewicz Research Network – Krakow Institute of Technology, Poland.

## References

[1]    T.M. Tritt, M.A. Subramanian, Thermoelectric Materials, Phenomena, and Applications: A Bird's Eye View, MRS Bull. 31 (2006) 188–198. https://doi.org/10.1557/mrs2006.44.

[2]    H. Ohta, K. Sugiura, K. Koumoto, Recent progress in oxide thermoelectric materials: P-type Ca 3Co4O9 and n-Type SrTiO3-, Inorg. Chem. 47 (2008) 8429–8436. https://doi.org/10.1021/ic800644x.

[3]    J.W. Fergus, Oxide materials for high temperature thermoelectric energy conversion, J. Eur. Ceram. Soc. 32 (2012) 525–540. https://doi.org/10.1016/j.jeurceramsoc.2011.10.007.

[4]    M. Dutta, T. Ghosh, K. Biswas, Electronic structure modulation strategies in high-performance thermoelectrics, APL Mater. 8 (2020) 040910. https://doi.org/10.1063/5.0002129.

[5]    R. Stern, T. Wang, J. Carrete, N. Mingo, G.K.H. Madsen, Influence of point defects on the thermal conductivity in FeSi, Phys. Rev. B. 97 (2018) 195201. https://doi.org/10.1103/PhysRevB.97.195201.

[6]    D.A. Dalton, W.P. Hsieh, G.T. Hohensee, D.G. Cahill, A.F. Goncharov, Effect of mass disorder on the lattice thermal conductivity of MgO periclase under pressure, Sci. Rep. 3 (2013) 1–5. https://doi.org/10.1038/srep02400.

[7]    D. Sivaprahasam, S.B. Chandrasekhar, S. Kashyap, A. Kumar, R. Gopalan, Thermal conductivity of nanostructured Fe0.04Co0.96Sb3 skutterudite, Mater. Lett. 252 (2019) 231–234. https://doi.org/10.1016/j.matlet.2019.05.140.

[8]    P.C. Sreeparvathy, V. Kanchana, Novel natural super-lattice materials with low thermal conductivity for thermoelectric applications: A first principles study, J. Phys. Chem. Solids. 111

(2017) 54–62. https://doi.org/10.1016/j.jpcs.2017.07.009.

[9]     A. Kumar, R. Kumar, D.K. Satapathy, Bi2Se3-PVDF composite: A flexible thermoelectric system, Phys. B Condens. Matter. 593 (2020) 412275. https://doi.org/10.1016/j.physb.2020.412275.

[10]    A.G. Every, Y. Tzou, D.P.H. Hasselman, R. Raj, The effect of particle size on the thermal conductivity of ZnS/diamond composites, Acta Metall. Mater. 40 (1992) 123–129. https://doi.org/10.1016/0956-7151(92)90205-S.

[11]    A. Kumar, D. Sivaprahsam, A.D. Thakur, Improvement of thermoelectric properties of lanthanum cobaltate by Sr and Mn co-substitution, J. Alloys Compd. 735 (2018) 1787–1791. https://doi.org/10.1016/j.jallcom.2017.11.334.

[12]    J. Androulakis, P. Migiakis, J. Giapintzakis, La0.95Sr0.05CoO3: An efficient room-temperature thermoelectric oxide, Appl. Phys. Lett. 84 (2004) 1099–1101. https://doi.org/10.1063/1.1647686.

[13]    A. Kumar, C. V Tomy, A.D. Thakur, Magnetothermopower, magnetoresistance and magnetothermal conductivity in La 0.95 Sr 0.05 Co 1− x Mn x O 3 (0.00 ≤ x ≤ 1.00), Mater. Res. Express. 5 (2018) 086110. https://doi.org/10.1088/2053-1591/aad44c.

[14]    J. Pei, G. Chen, D.Q. Lu, P.S. Liu, N. Zhou, Synthesis and high temperature thermoelectric properties of Ca3.0-x-yNdxNayCo4O9+δ, Solid State Commun. 146 (2008) 283–286. https://doi.org/10.1016/j.ssc.2008.03.012.

[15]    D. Kenfaui, D. Chateigner, M. Gomina, J.G. Noudem, Texture, mechanical and thermoelectric properties of Ca3Co4O9 ceramics, J. Alloys Compd. 490 (2010) 472–479. https://doi.org/10.1016/j.jallcom.2009.10.048.

[16]    W. Koshibae, K. Tsutsui, S. Maekawa, Thermopower in cobalt oxides, Phys. Rev. B. 62 (2000) 6869–6872. https://doi.org/10.1103/PhysRevB.62.6869.

[17]    I. Terasaki, Y. Sasago, K. Uchinokura, Large thermoelectric power in NaCo2O4 single crystals, Phys. Rev. B - Condens. Matter. 56 (1997) R12685. https://doi.org/https://doi.org/10.1103/PhysRevB.56.R12685.

[18]    K. Mydeen, P. Mandal, D. Prabhakaran, C.Q. Jin, Pressure- and temperature-induced spin-state transition in single-crystalline La1-x Srx CoO3 (x=0.10 and 0.33), Phys. Rev. B - Condens. Matter Mater. Phys. 80 (2009) 1–6. https://doi.org/10.1103/PhysRevB.80.014421.

[19]    K. Koumoto, Y. Wang, R. Zhang, A. Kosuga, R. Funahashi, Oxide thermoelectric materials: A nanostructuring approach, Annu. Rev. Mater. Res. 40 (2010) 363–394. https://doi.org/10.1146/annurev-matsci-070909-104521.

[20]    Z. Shi, F. Gao, J. Xu, J. Zhu, Y. Zhang, T. Gao, M. Qin, M. Reece, H. Yan, Two–step processing of thermoelectric (Ca 0.9 Ag 0.1 ) 3 Co 4 O 9 /nano–sized Ag composites with high ZT, J. Eur. Ceram. Soc. 39 (2019) 3088–3093. https://doi.org/10.1016/j.jeurceramsoc.2019.04.004.

[21]    A. Kumar, K. Kumari, S.J. Ray, A.D. Thakur, Graphene mediated resistive switching and thermoelectric behavior in lanthanum cobaltate, J. Appl. Phys. 127 (2020) 235103. https://doi.org/10.1063/5.0009666.

[22]    A. Kumar, K. Kumari, B. Jayachandran, D. Sivaprahsam, A.D. Thakur, Thermoelectric properties of (1-x)LaCoO3.xLa0.7Sr0.3MnO3 composite, J. Alloys Compd. 749 (2018) 1092–1097. https://doi.org/10.1016/j.jallcom.2018.03.347.

[23]    A. Kumar, K. Kumari, B. Jayachandran, D. Sivaprahsam, A.D. Thakur, Thermoelectric

Properties of $(1 − x)$LaCoO 3 .$( x )$La 0.95 Sr 0.05 CoO 3 composite, Mater. Res. Express. 6 (2019) 055502. https://doi.org/10.1088/2053-1591/aade73.

[24] K. Xu, H. Liu, Introducing Ta2O5 nanoparticles into Ca3Co4O9 matrix for increased thermoelectric property through phonon scattering, Ceram. Int. 46 (2020) 25783–25788. https://doi.org/10.1016/j.ceramint.2020.07.057.

[25] P. Jood, R.J. Mehta, Y. Zhang, G. Peleckis, X. Wang, R.W. Siegel, T. Borca-Tasciuc, S.X. Dou, G. Ramanath, Al-doped zinc oxide nanocomposites with enhanced thermoelectric properties, Nano Lett. 11 (2011) 4337–4342. https://doi.org/10.1021/nl202439h.

[26] N. Wang, H. He, X. Li, L. Han, C. Zhang, Enhanced thermoelectric properties of Nb-doped SrTiO3 polycrystalline ceramic by titanate nanotube addition, J. Alloys Compd. 506 (2010) 293–296. https://doi.org/10.1016/j.jallcom.2010.06.195.

[27] M. Wolf, K. Menekse, A. Mundstock, R. Hinterding, F. Nietschke, O. Oeckler, A. Feldhoff, Low Thermal Conductivity in Thermoelectric Oxide-Based Multiphase Composites, J. Electron. Mater. 48 (2019) 7551–7561. https://doi.org/10.1007/s11664-019-07555-2.

[28] D.P.H. Hasselman, K.Y. Donaldson, A.L. Geiger, Effect of Reinforcement Particle Size on the Thermal Conductivity of a Particulate-Silicon Carbide-Reinforced Aluminum Matrix Composite, J. Am. Ceram. Soc. 75 (1992) 3137–3140. https://doi.org/10.1111/j.1151-2916.1992.tb04400.x.

[29] X. Liang, M. Baram, D.R. Clarke, Thermal (Kapitza) resistance of interfaces in compositional dependent ZnO-In 2 O 3 superlattices, Appl. Phys. Lett. 102 (2013) 223903. https://doi.org/10.1063/1.4809784.

[30] K. Pietrak, T.S. Wiśniewski, M. Kubiś, Application of flash method in the measurements of interfacial thermal resistance in layered and particulate composite materials, Thermochim. Acta. 654 (2017) 54–64. https://doi.org/10.1016/j.tca.2017.05.007.

[31] A. Kumar, A. Kosonowski, P. Wyzga, K.T. Wojciechowski, Effective thermal conductivity of SrBi4Ti4O15-La0.7Sr0.3MnO3 oxide composite: Role of particle size and interface thermal resistance, J. Eur. Ceram. Soc. 41 (2021) 451–458. https://doi.org/10.1016/j.jeurceramsoc.2020.08.069.

[32] A. Kumar, M. Battabyal, A. Chauhan, G. Suresh, R. Gopalan, N. V Ravi kumar, D.K. Satapathy, Charge transport mechanism and thermoelectric behavior in Te:(PEDOT:PSS) polymer composites, Mater. Res. Express. 6 (2019) 115302. https://doi.org/10.1088/2053-1591/ab43a7.

[33] K. Asai, A. Yoneda, O. Yokokura, J.M. Tranquada, G. Shirane, K. Kohn, Two Spin-State Transitions in LaCoO3, J. Phys. Soc. Japan. 67 (1998) 290–296. https://doi.org/10.1143/JPSJ.67.290.

[34] K. Asai, O. Yokokura, N. Nishimori, H. Chou, J.M. Tranquada, G. Shirane, S. Higuchi, Y. Okajima, K. Kohn, Neutron-scattering study of the spin-state transition and magnetic correlations in La1-xSrxCoO3 (x=0 and 0.08), Phys. Rev. B. 50 (1994) 3025–3032. https://doi.org/10.1103/PhysRevB.50.3025.

[35] C. KITTEL, Introduction to Solid State Physics (7th ed.). Wiley-India. ISBN 81-265-1045-5., 1995.

[36] K. Chen, W. Xiao, Z. Li, J. Wu, K. Hong, X. Ruan, Effect of Graphene and Carbon Nanotubes on the Thermal Conductivity of WC–Co Cemented Carbide, Metals (Basel). 9 (2019) 377. https://doi.org/10.3390/met9030377.

[37]    H.S. Kim, Z.M. Gibbs, Y. Tang, H. Wang, G.J. Snyder, Characterization of Lorenz number with Seebeck coefficient measurement, APL Mater. 3 (2015) 041506. https://doi.org/10.1063/1.4908244.

[38]    E.T. Swartz, R.O. Pohl, Thermal boundary resistance, Rev. Mod. Phys. 61 (1989) 605–668. https://doi.org/10.1103/RevModPhys.61.605.

[39]    D.A.G. Bruggeman, Bereclmung Verschiedener Physikalischer Konnstanten Won Heterogenen Substanxen I. Dlelehtrizitatskonstanten und Leitfiihigkeiten der Misciikiirper aus Isotropen Substanzen, Ann. Phys. 416 (1935) 636–664.

[40]    K. Iwasaki, T. Ito, T. Nagasaki, Y. Arita, M. Yoshino, T. Matsui, Thermoelectric properties of polycrystalline La1-xSrxCoO3, J. Solid State Chem. 181 (2008) 3145–3150. https://doi.org/10.1016/j.jssc.2008.08.017.

[41]    T. He, J. Chen, T.G. Calvarese, M.A. Subramanian, Thermoelectric properties of La 1−x A x CoO 3 (A = Pb, Na), Solid State Sci. 8 (2006) 467–469. https://doi.org/10.1016/j.solidstatesciences.2006.01.002.

[42]    F. Li, J.F. Li, Effect of Ni substitution on electrical and thermoelectric properties of LaCoO3 ceramics, Ceram. Int. 37 (2011) 105–110. https://doi.org/10.1016/j.ceramint.2010.08.024.